\documentclass[twocolumn,twoside,slac]{revtex4}
\usepackage{graphicx}
\usepackage{fancyhdr}
\usepackage{comment}
\begin{document}


\title{Next-Generation EU DataGrid Data Management Services}

%

\author{Diana Bosio, James Casey, Akos Frohner, Leanne Guy,  Peter
  Kunszt, Erwin Laure, Sophie Lemaitre, Levi Lucio, Heinz Stockinger, Kurt Stockinger}

\affiliation{CERN, European Organization for Nuclear Research, CH-1211
  Geneva 23, Switzerland}

\author{William Bell, David Cameron, Gavin McCance, Paul Millar}
\affiliation{University of Glasgow, Glasgow, G12 8QQ, Scotland}

\author{Joni Hahkala, Niklas Karlsson, Ville Nenonen, Mika Silander}
\affiliation{Helsinki Institute of Physics, P.O. Box 64, 00014 University of
Helsinki, Finland}

\author{Olle Mulmo, Gian-Luca Volpato}
\affiliation{Swedish Research Council, SE-103 78 Stockholm, Sweden}

\author{Giuseppe Andronico}
\affiliation{INFN Catania, Via S. Sofia, 64, I-95123 Catania, Italy}

\author{Federico DiCarlo}
\affiliation{INFN Roma, P.le Aldo Moro, 2,  I-00185 Roma, Italy}

\author{Livio Salconi}
\affiliation{INFN Pisa, via F. Buonarroti 2, I-56127 Pisa, Italy}

\author{Andrea Domenici}
\affiliation{DIIEIT, via Diotisalvi, 2, I-56122 Pisa, Italy}

\author{Ruben Carvajal-Schiaffino, Floriano Zini}
\affiliation{ITC-irst, via Sommarive 18, 38050 Povo, Trento, Italy}

\begin{abstract}
We describe the architecture and initial implementation of the 
next-generation of Grid Data Management Middleware in the EU DataGrid 
(EDG) project.

The new architecture stems from our experience together with the user 
requirements gathered during the two years of running our initial set of 
Grid Data Management Services. All of our new services are based on the 
Web Service technology paradigm, very much in line with the emerging 
Open Grid Services Architecture (OGSA). We have modularized our 
components and invested a great amount of effort in developing  secure, 
extensible and robust services, starting from the design but also using a 
streamlined build and testing framework.

Our service components are: Replica Location Service, Replica Metadata 
Service, Replica Optimization Service, Replica Subscription and 
high-level replica management. The service security infrastructure is 
fully GSI-enabled, hence compatible with the existing Globus Toolkit 
2-based services; moreover, it allows for fine-grained authorization 
mechanisms that can be adjusted depending on the service semantics.
\end{abstract}

\maketitle

\thispagestyle{fancy}

\section{Introduction}

The EU DataGrid project~\cite{EDG} (also referred to as EDG in this article) is
now in its third and final year. Within the data management
work package we have developed a second generation of data management
services that will be deployed in EDG release 2.x. Our first
generation replication tools (GDMP, edg-replica-manager etc.) provided
a very good base and input, which we reported on 
in~\cite{Sto01,Sto03}. The experience we gained in the first generation of
tools (mainly written in C++), is directly used in the second
generation of data management services that are based on web service
technologies and mainly implemented in Java.

The basic design concepts in the second generation services are as
follows:

\begin{itemize}

\item Modularity: 

  The design needs to be modular and allow for easy
  plug-ins and future extensions. 

  In addition, we should use generally
  agreed standards and do not rely on vendor specific solutions.

\item Evolution: 

  Since OGSA is an upcoming standard that is most
  likely to be adapted by several Grid services in the future, the
  design should allow for an easy adoption of the OGSA concept. It is
  also advisable to use a similar technology. 

  In addition, the design
  should be independent of the underlying operating system as well as
  relational database managements system that are used by our services.

\end{itemize}

Having implemented the first generation tools mainly in C++, the
technology choices for the second generation services presented in this
article are as follows:

\begin{itemize}

\item Java based servers are used that host web services (mainly
  Jakarta's Tomcat as well as Oracle 9iAS for certain applications).

\item Interface definitions in WSDL

\item Client stubs for several programming languages (Java, C/C++)
  through SOAP using AXIS for Java and gSOAP for C++ interfaces.

\item Persistent service data is stored in a relational database
  management system. We mainly use MySQL for general services that
  require open source technology and Oracle for more robust services.

\end{itemize}

The entire set of data management services consists of the following
parts:

\begin{itemize}

\item {\bf Replication service framework}: This service framework is
  the main part of our data management services and is described in
  detail in Section~\ref{section:replication}. It basically consists
  of an overall replica management system that uses several other
  services such as  the Replica Location Service, Replica Optimization
  service etc.

\item {\bf SQL Database Service (Spitfire)}: Spitfire provides a 
  means to access relational databases from the Grid.

\item {\bf Java Security Package}: All of our services have very
  strict security requirements. The Java security package provides
  tools that can be used in Grid services such as our replication
  services.

\end{itemize}

All these components are discussed in detail in the following sections
and thus also outline the paper organization.

\section{Replication Service Framework 'Reptor'}
\label{section:replication}

In the following section we first give an architectural overview of
the entire replication framework and then discuss individual
services (Replica Location Service, Replica Optimization Service etc.) in
more detail.

\subsection{General Overview of Replication Architecture}

Figure \ref{figure:RMSDesign} presents the user's perspective of the
main components of a replica management system for which we have given
the code-name `Reptor'. This design, which first was discussed
in~\cite{Reptor}, represents an evolution of the original design
presented in~\cite{EDG2.2, Hoschek00}. Several of the components have
already been implemented and tested in EDG (see shaded components) whereas
others (in white) are still in the design phase and might be
implemented in the future.

Reptor has been realized as a modular system that provides easy plugability of third
party components. Reptor defines the minimal interface third party
components have to provide. According to this design the entire framework is provided by
the {\bf Replica Management Service} which acts as a logical single
entry point to the system and interacts with the other components of
the systems as follows:

\begin{figure*}[htb]
\centering
\includegraphics[width=130mm]{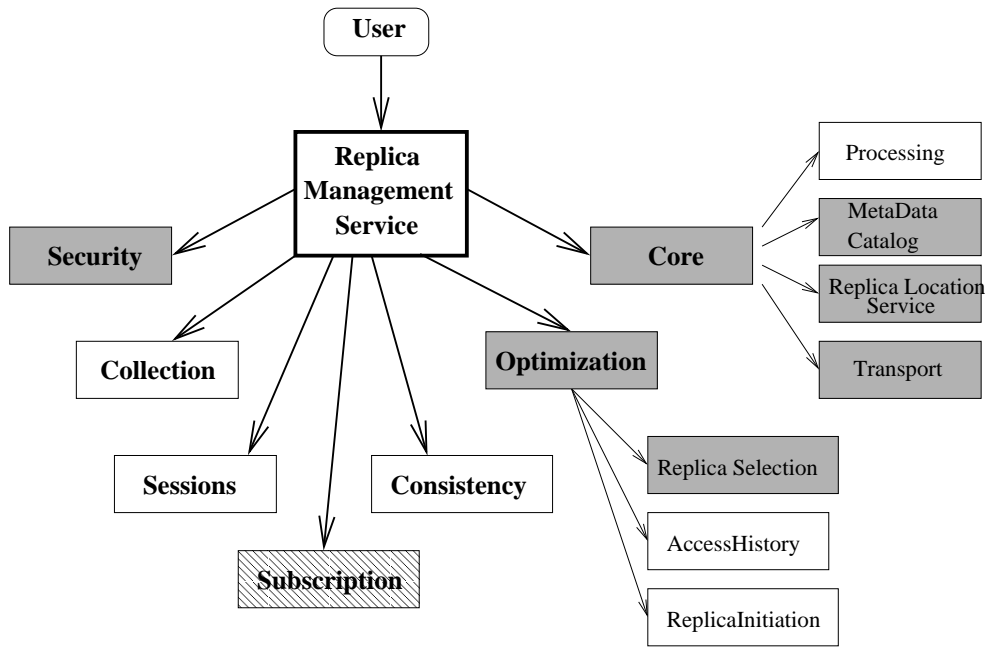}
\caption{Reptor's main design components.} 
\label{figure:RMSDesign}
\end{figure*}

\begin{itemize}
\item The {\bf Core} module provides the main functionality of replica
management, namely replica creation, deletion, and cataloging by
interacting with third party modules such as transport and replica and
metadata catalog services.

\item The goal of the {\bf Optimization} component (implemented as a
  service) is to minimize file access times by pointing access
  requests to appropriate replicas and pro-actively replicating
  frequently used files based on gathered access statistics.

\item The {\bf Security} module manages the required user
authentication and authorization, in particular, issues pertaining to
whether a user is allowed to create, delete, read, and write a file. 

\item {\bf Collections} are defined as sets of logical filenames and
  other collections.

\item The {\bf Consistency} module maintains consistency
between all   replicas of a given file, as well as between the meta
information stored in the various catalogs.

\item The {\bf Session} component provides generic check-pointing,
restart, and rollback mechanisms to add fault tolerance to the
system.

\item The {\bf Subscription} service allows for a publish-subscribe
  model for replica creation.

\end{itemize}

We decided to implement the Replica Management
Service and the core module functionality on the client side in the
Replica Manager Client, henceforth referred to as the {\bf Replica Manager}.
The other subservices and APIs are modules and
services in their own right, allowing for a multitude of deployment
scenarios in a distributed environment. 

One advantage of such a design is that if a subservice is unavailable,
the Replica Manager can still provide all the functionality
that does not make use of that particular service. Also, critical
service components may have more than one instance to provide a higher
level of availability and to avoid service bottlenecks.

A detailed description of the implemented components and services can
be found in the following subsections as well as in the original
design in~\cite{Reptor}.


\subsection{Interaction with Services}

The Replica Manager needs to interact
with many external services as well as internal ones,
such as the the Information Service and transport mechanisms
like GridFTP servers~\cite{All02}. Most of the components
required by the Replica Manager are independent services, hence appropriate client stubs
satisfying the interface need to be provided by the service. By
means of configuration files the actual component to be used can be
specified and Java dynamic class loading features are exploited for
making them available at execution time.

To date, the Replica Manager has been  tested using the following components:

\begin{itemize}

\item {\em Replica Location Service (RLS)}~\cite{Giggle}: 
  used for locating replicas in the Grid and assigning physical file
  names.  

\item {\em Replica Metadata Catalog (RMC)}: used for querying and 
assigning logical file names.

\item {\em Replica Optimization Service (ROS)}: used for locating
the best replica to access.

\item {\em R-GMA}: an information service provided by EDG: The Replica Manager
  uses R-GMA to  obtain information about Storage and Computing
  Elements~\cite{Reptor}.

\item {\em Globus C based libraries as well as CoG}~\cite{cog}
  providing GridFTP transport functionality.

\item The {\em EDG network monitoring services}: EDG (in particular
  WP7) provides these
  services to obtain statistics and network characteristics.

%

\end{itemize}

\begin{figure*}[htb]
\centering
\includegraphics[width=130mm]{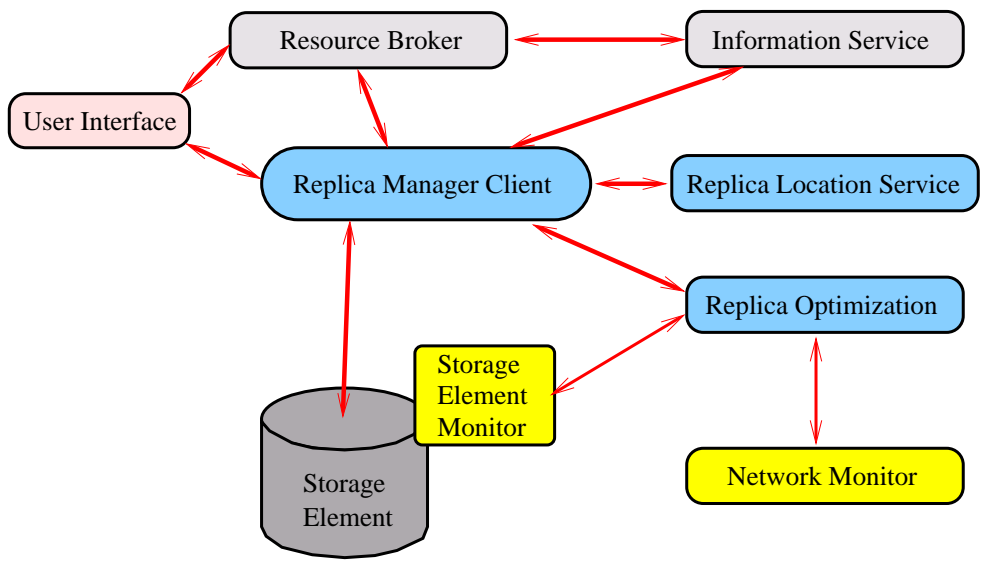}
\caption{Interaction of Replica Manager with other Grid components.}
\label{figure:ReplicaManagerInteractions}
\end{figure*}

The implementation is mainly done using the Java J2EE framework and
associated web service technologies (the Apache Tomcat servlet
container, Jakarta Axis , etc.). In more detail, we use client/server
architectures making SOAP Remote Procedure Call (RPC) over HTTPS. The basic component
interaction is given in Figure~\ref{figure:ReplicaManagerInteractions}
and will also explained in a few more details in the following sub
sections. For more details on web service choices refer to
Section~\ref{sec:web-sevice}.

For the user, the main entry point to the Replication Services is
through the client interface that is provided via a Java API as well as
a command line interface, the {\tt edg-replica-manager} module. For
each of the main components in Figure~\ref{figure:RMSDesign}, the
Reptor framework provides the necessary interface. For instance, the
functionality of the core module includes mainly the file copy and
cataloging process and is handled in the client library with the
respective calls to the Transport and Replica Catalog modules.

\subsection{Replica Location Service (RLS)}

The Replica Location Service (RLS) is the service responsible for
maintaining a (possibly distributed) catalog of files registered in the Grid
infrastructure. For each file there may exist several replicas. This
is due to the need for geographically distributed copies of the same
file, so that accesses from different points of the globe may be
optimized (see section on the Replica Optimization
Service). Obviously, one needs to keep track of the scattered
replicas, so that they can be located and consistently updated.

As such, the RLS is designed to store one-to-many relationships
between (Grid Unique Identifiers (GUIDs) and Physical File Names
(PFNS). Since many replicas of the same file may coexist (with different
PFNs) we identify them as being replicas of the same file by assigning
to them the same unique identifier (the GUID).

The RLS architecture encompasses two logical components - the LRC
(Local Replica Catalog) and the RLI (Replica Location Index). 
The LRC stores the mappings between GUIDs and PFNs on a per-site
basis whereas the RLI stores information on where mappings
exist for a given GUID. In this way, it is possible to split the
search for replicas of a given file in two steps: in the first one the
RLI is consulted in order to determine which LRCs contain mappings
for a given GUID; in the second one, the specific LRCs are consulted in
order to find the PFNs one is interested in.

It is however worth mentioning that the LRC is implemented to work in
standalone mode, meaning that it can act as a full RLS on its own if
such a deployment architecture is necessary.  When working in
conjunction with one (or several) RLIs, the LRC provides periodic
updates of the GUIDs it holds mappings for. These updates consist of
bloom filter objects, which are a very compact form of representing a
set, in order to support membership queries~\cite{bloom}.

The RLS currently has two possible database backend deployment
possibilities: MySQL and Oracle9i.

\subsection{Replica Metadata Catalog Service (RMC)}

Despite the fact that the RLS already provides the necessary
functionality for application clients, the GUID unique identifiers
are difficult to read and remember.  The
Replica Metadata Catalog (RMC) can be considered as another layer of
indirection on top of the RLS that provides mappings between Logical
File Names (LFNs) and GUIDs. The LFNs are user defined aliases for
GUIDs - many LFNs may exist for one GUID.

Furthermore, the RMC is also capable of holding metadata about the
original physical file represented by the GUID (e.g. size, date of
creation, owner). It is also possible for the user to define specific
metadata and attach it to a GUID or to an LFN. The purpose of this
mechanism is to provide to users and applications a way of querying
the file catalog based on a wide range of attributes. The possibility
of gathering LFNs as collections and manipulating these collections as
a whole has already been envisaged, but is not yet implemented.

As for the RLS, the RMC supports MySQL and Oracle9i as database backends.

\subsection{Replica Optimization Service (ROS)}
The goal of the optimization service is to select the best replica
with respect to network and storage access latencies. It is
implemented as a light-weight web service that gathers information from the EDG
network monitoring service and the EDG storage element service about
the respective data access latencies.

In~\cite{BCC+02a} we defined the APIs \texttt{getNetworkCosts} and \texttt{getSECosts}
for interactions of the Replica Manager with the Network Monitoring
and the Storage Element Monitor. These two components monitor the network
traffic and the access traffic to the storage device respectively and
calculate the expected transfer time of a given file with a specific
size.

In the EU DataGrid Project, Grid resources are managed by the meta
scheduler of WP1, the Resource Broker~\cite{EDG-WP1}. One of the goals
of the Resource Broker is to decide on which Computing Element the
jobs should be run such that the throughput of all jobs is
maximized. Assuming highly data intensive jobs, a typical optimization
strategy could be to select the least loaded resource with the maximum
amount of locally avaliable data. In~\cite{BCC+02a} we introduced the
Replica Manager API
\texttt{getAccessCost} that returns the access costs of a specific job
for each candidate Computing Element. The Resource Broker can then
take this information provided by the Replica Manager to schedule each
job to its optimal resources.

The interaction of the Replica Manager with the Resource Broker, the
Network Monitor and the Storage Element Monitor is depicted in Figure
\ref{figure:ReplicaManagerInteractions}.

\subsection{Replica Subscription Service} 

The Replica Subscription Service (RSS) provides automatic
replication based on a subscription model. The basic design is based
on our first generation replication tool GDMP (Grid Data Mirroring
Package)~\cite{Sto03}.

\section{SQL Database Service: Spitfire}

Spitfire~\cite{spitfire} provides a means to access relational
databases from the Grid. This service has been provided by our work
package for some time and was our first service that used the web
service paradigm. Thus, we give more details about its implementation in
Section~\ref{sec:web-sevice} since many of the technology choices for
the replication services explained in the previous section are based
on choices also made for Spitfire.

\subsection{Spitfire Overview}

The SQL Database service (named Spitfire) permits convenient and
secure storage, retrieval and querying of data held in any local or
remote RDBMS.  The service is optimized for metadata storage.  The
primary SQL Database service has been re-architected into a standard
web service.  This provides a platform and language independent way of
accessing the information held by the service. The service exposes a
standard interface in WSDL format, from which client stubs can be
built in most common programming languages, allowing a user
application to invoke the remote service directly.  The interface
provides the common SQL operations to work with the data.  Pre-built
client stubs exist for the Java, C and C++ programming languages.  The
service itself has been tested with the MySQL and Oracle databases.

The earlier SQL Database service was primarily accessed  via  a  web  browser
(or  command   line)   using   pre-defined   server-side   templates.   This
functionality, while less flexible than the  full  web  services  interface,
was found to be very useful for web portals, providing a  standardized  view
of the data. It has therefore been retained and re-factored into a  separate
SQL Database browser module.

\subsection{Component Description and Details about Web Service Design}
\label{sec:web-sevice}

There are three main components to  the  SQL Database  service:  the  primary
server component,  the  client(s)  component,  and  the  browser  component.
Applications that  have  been  linked  to  the  SQL Database  client  library
communicate to a remote instance of the server. This server is put in  front
of a RDBMS (e.g. MySQL), and securely  mediates  all  Grid  access  to  that
database. The browser is a standalone web portal  that  is  also  placed  in
front of a RDBMS.

The server is a fully compliant web service implemented in Java. It
runs on Apache Axis inside a Java servlet engine (currently we use the
Java reference servlet engine, Tomcat, from the Apache Jakarta
project).  The service mediates the access to a RDBMS that must be
installed independently from the service.  The service is reasonably
non-intrusive, and can be installed in front of a pre-existing RDBMS.
The local database administrator retains full control of the database
back-end, with only limited administration rights being exposed to
properly authorized grid users.

The web services client, at its most  basic,  consists  of  a  WSDL  service
description  that  describes  fully   the   interface.   Using   this   WSDL
description, client stubs can be generated automatically in the  programming
language of choice. We provide pre-built client stubs for the  Java,  C  and
C++ programming languages. These are packaged as Java JAR files  and  static
libraries for Java and C/C++ respectively.

The browser component is a server side component that provides
web-based access to the RDBMS.  It provides the functionality of the
previous version of the SQL Database service.  This service does not
depend on the other components and can be used from any web browser.
The browser component is implemented as a Java servlet.  In the case
where it is installed together with the primary service, it is envisaged
that both services will be installed inside the same servlet engine.

The design of the primary service is similar to that of  the  prototype  Remote
Procedure Call GridDataService standard discussed in~\cite{dais}, and  indeed,
influenced the design of the standard. It is expected that  the  SQL Database
service will eventually evolve into a prototype implementation  of  the  RPC
part  of  this  GGF  standard.  However,  to  maximise  the  usability   and
portability of the service,  we  chose  to  implement  it  as  a  plain  web
service, rather than just an OGSA service. The architecture of  the  service
has been designed  so  that  it  will  be  trivial  to  implement  the  OGSA
specification at a later date.

The communication between the client  and  server  components  is  over  the
HTTP(S) protocol. This maximises the portability of the service, since  this
protocol has many pre-existing applications that have  been  heavily  tested
and are now very robust. The data format is  XML,  with  the  request  being
wrapped  using  standard  SOAP  Remote  Procedure  Call.  The  interface  is
designed around the  SQL  query  language.  The  communication  between  the
user's web  browser  and  the  SQL Database  Browser  service  is  also  over
HTTP(S).

The server and browser components (and parts of the Java client  stub)  make
use of  the  common  Java  Security  module  as described in
Section~\ref{sec:security}.  The  secure connection is made over HTTPS
(HTTP with SSL or TLS).

Both the server and browser have a service certificate (they can
optionally make use of the system's host certificate), signed by an
appropriate CA, which they can use to authenticate themselves to the
client.  The client uses their GSI proxy to authenticate themselves to
the service. The user of the browser service should load their GSI
certificate into the web browser, which will then use this to
authenticate the user to the browser.

A basic authorisation scheme is  defined  by  default  for  the  SQL Database
service, providing  administrative  and  standard  user  functionality.  The
authorisation is performed using the subject name of the user's  certificate
(or a regular expression matching it). The  service  administrator  can  define  a
more  complex  authorisation  scheme  if  necessary,  as  described  in  the
security module documentation.

\section{ Security }
\label{sec:security}

The EDG Java security package covers two main security areas, 
authentication authorization.
Authentication assures that the entity (user,
service or server) at the other end of the connection is who it claims
to be. Authorization decides what the entity is allowed to
do.

The aim in the security package is always to make the software as
flexible as possible and to take into account the needs of both EDG
and industry to make the software usable everywhere. To this end there
has been some research into similarities and possibilities for
cooperation with for example Liberty Alliance, which is a consortium
developing standards and solutions for federated identity for web
based authentication, authorization and payment.

\subsection{ Authentication }

The authentication mechanism is an extension of the normal Java SSL
authentication mechanism. The mutual authentication in SSL happens by
exchanging public certificates that are signed by trusted certificate
authorities (CA). The user and the server prove that they are the owners of the
certificate by proving in cryptographic means that they have the
private key that matches with the certificate.

In Grids the authentication is done using GSI proxy certificates that are
derived from the user certificate. This proxy certificate comes close to
fulfilling the {\em PKIX}~\cite{PKIX} requirement for valid certificate 
chain, but does not
fully follow the standard. This causes the SSL handshake to fail in the
conforming mechanisms. For the GSI proxy authentication to work the SSL
implementation has to be nonstandard or needs to be changed to accept them.

The EDG Java security package extends the Java SSL package. It

\begin{itemize}
\item accepts the GSI proxies as the authentication method
\item supports GSI proxy loading with periodical reloading
\item supports OpenSSL certificate-private key pair loading
\item supports CRLs with periodical reloading
\item integrates with Tomcat
\item integrates with Jakarta Axis SOAP framework
\end{itemize}

The GSI proxy support is done by finding the user certificate and making
special allowances and restrictions to the following proxy certificates. The
allowance is that the proxy certificate does not have to be signed by a CA.
The restriction is that the distinguished name (DN) of the proxy certificate
has to start with the DN of the user certificate (e.g. `C=CH, O=cern, 
CN=John Doe'). This way the user cannot pretend to be someone else by making
a proxy with DN `C=CH, O=cern, CN=Jane Doe'. The proxies are short lived, so
the program using the SSL connection may be running while the proxy is
updated. For this reason the user credentials (for example the proxy
certificate) can be made to be reloaded periodically.

OpenSSL saves the user credentials using two files, one for the user
certificate and the other for the private key. With the EDG Java security
package these credentials can be loaded easily.

The CAs periodically release lists of revoked certificates in a certificate
revocation list (CRL). The EDG Java security package supports this CRL
mechanism and even if the program using the package is running, these lists
can be periodically and automatically reloaded into the program by setting
the reload interval.

The integration to Jakarta Tomcat (a Java web server and servlet
container) is done with an interface class and to use it only the
Jakarta Tomcat configuration file has to be set up accordingly.

The Jakarta Axis SOAP framework provides an easy way to change the
underlying SSL socket implementation on the client side. Only a simple
interface class was needed and to turn it on a system variable has to
be set while calling the Java program. In the server side the
integration was even simpler as Axis runs on top of 
Tomcat and Tomcat can be set up as above.

Due to issues of performance, many of the services described in this
document have equivalent clients written in C++. To this end, there
are several C++ SOAP clients that have been written based on the gSOAP
library. In order to provide the same authentication and authorization
functionality as in the corresponding Java SOAP clients, an
accompanying C library is being developed for gSOAP.  When ready, it
is to provide support for mutual authentication between SOAP clients
and SOAP servers, support for the coarse-grained authorization as
implemented in the server end by the Authorization Manager (described 
below) and verification of both standard X509 and GSI style server and 
server proxy certificates.

\subsection{ Coarse grained authorization }

The EDG Java security package only implements the coarse grained
authorization. The coarse grained authorization decision is made in the 
server before the actual call to the service and can make decisions such as
`what kind of access does this user have to that database table' or `what
kind of access does this user have to the file system'. The fine grained
authorization that answers the question `what kind of access does this user
have to this file' can only be handled inside the service, because the
actual file to access is only known during the execution of the service.
The authorization mechanism is positioned in the server before the service.

In the EDG Java security package the authorization is implemented as role
based authorization. Currently the authorization is done in the server end
and the server authorizes the user, but there are plans to do mutual
authorization where also the client end checks that the server end is
authorized to perform the service or to save the data. The mutual
authorization is especially important in the medical field where the medical
data can only be stored in trusted servers.

The role based authorization happens in two stages, first the system checks
that the user can play the role he requested (or if there is a default role
defined for him). The role the user is authorized to play is then mapped to a
service specific attribute. The role definitions can be the same in all the
services in the (virtual) organization, but the mapping from the role to
the attribute is service specific. The service specific attribute can be for
example a user id for file system access of database connection id with
preconfigured access rights. If either step fails, the user is not
authorized to access the service using the role he requested.

There are two modules to interface to the information flow
between the client and the service; one for normal HTTP web traffic
and the other for SOAP web services. The authorization mechanism can
attach to other information flows by writing a simple interface
module for them.

In a similar fashion the authorization information that is used to make
the authorization decisions can be stored in several ways. For simple
and small installation and for testing purposes the information can be
a simple XML file. For larger installations the information can be
stored into a database and when using the Globus tools to distribute
the authorization information, the data is stored in a text file that
is called the gridmap file. For each of these stores there is a module to
handle the specifics of that store and to add a new way to store the
authorization information. Only a interface module needs to be written.
When the virtual organization membership service (VOMS) is used the
information provided by the VOMS server can be used for the authorization
decisions and all the information from the VOMS is parsed and forwarded to
the service.

\subsection{ Administration web interface }

The authorization information usually ends up being rather complex,
and maintaining that manually would be difficult, so a web based
administration interface was created. This helps to understand the
authorization configuration, eases the remote management and by making
management easier improves the security.

\section{Conclusions}

The second generation of our data management services has been
designed and implemented based on the web service paradigm. In this
way, we have a flexible and extensible service framework and are thus
prepared to follow the general trend of the upcoming OGSA standard
that is based on web service technology. Since interoperability of
services seems to be a key feature in the upcoming years, we believe
that our approach used in the second generation of data management is
compatible with the need for service interoperability in a rapidly changing
Grid environment.

Our design choices have been as follows: we aim for supporting robust,
highly available commercial products (like Oracle/DB and Oracle/Application Server) as well as standard open source technology (MySQL, Tomcat,
etc.).

The first experience in using the new generation of services shows that
basic performance expectations are met. During this year, the
services will be deployed on the EDG testbed (and possibly others):
this will show the strength and the weaknesses of the services.

\begin{acknowledgments}
This work was partially funded by the European Commission program
IST-2000-25182 through the EU DataGrid Project.
\end{acknowledgments}


\begin{thebibliography}{9}

\bibitem{All02}
W. Allcock, J. Bester, J. Bresnahan, A. Chernevak,
I. Foster, C. Kesselman, S. Meder, V. Nefedova, D. Quesnal, S. Tuecke;
"Data Management and Transfer in High Performance Computational Grid
Environments." Parallel Computing, 2002.

\bibitem{BCC+02a} W. H. Bell, D. G. Cameron, L. Capozza, P. Millar,
  K. Stockinger, F. Zini, Design of a Replica Optimisation Framework,
  \emph{Technical Report, DataGrid-02-TED-021215}, Geneva,
  Switzerland, December 2002.

\bibitem{spitfire}
William Bell, Diana Bosio, Wolfgang Hoschek, Peter Kunszt, Gavin
McCance, and Mika Silander. ``Project Spitfire - Towards Grid Web
Service Databases''. Technical report, Global Grid Forum Informational
Document, GGF5, Edinburgh, Scotland, July 2002. 


\bibitem{Giggle} Ann Chervenak, Ewa Deelman, Ian Foster, Leanne Guy, Wolfgang Hoschek,
Adriana Iamnitchi, Carl Kesselman, Peter Kunszt, Matei Ripenu,
Bob Schwartzkopf, Heinz Stocking, Kurt Stockinger, Brian Tierney
, ``Giggle: A Framework for Constructing
  Scalable Replica Location Services'',Proceedings of SC2002 Conference,
  November 2002


\bibitem{EDG-WP1} DataGrid WP1, Definition of Architecture, Technical
  Plan and Evaluation Criteria for Scheduling, Resource Management,
  Security and Job Description, \emph{Technical Report, EU DataGrid
    Project. Deliverable D1.2}, September 2001.

\bibitem{EDG}
European DataGrid project (EDG): http://www.eu-datagrid.org


\bibitem{Reptor} L. Guy, P. Kunszt, E. Laure, H. Stockinger,
  K. Stockinger ``Replica Management in Data Grids'', Technical
  Report, GGF5 Working Draft, Edinburgh Scotland, July 2002


\bibitem{EDG2.2} Wolfgang Hoschek, Javier Jaen- Martinez, Peter
Kunszt, Ben Segal, Heinz Stockinger, Kurt Stockinger, Brian Tierney,
"Data Management (WP2) Architecture Report", EDG Deliverable 2.2,
http://edms.cern.ch/document/332390

\bibitem{Hoschek00}
Wolfgang Hoschek, Javier Jean-Martinez, Asad Samar, Heinz Stockinger,
Kurt Stockinger. Data Management in an International Data Grid
Project. \emph{1st IEEE/ACM International Workshop on Grid Computing
(Grid'2000)}. Bangalore, India, Dec 17-20, 2000.

\bibitem{PKIX} R. Housley et.al. ``Internet X.509 Public Key Infrastructure
Internet X.509 Public Key Infrastructure, RFC 3280, The Internet Society
April 2002, http://www.ietf.org/rfc/rfc3280.txt

\bibitem{dais} Amy Krause, Susan Malaika, Gavin McCance, James
Magowan, Norman W. Paton, Greg Riccardi ``Grid Database Service
Specification'', Global Grid Forum 6, Edinburgh, 2002.


\bibitem{cog}
Gregor von Laszewski, Ian Foster, Jarek Gawor, Peter Lane:
``A Java Commodity Grid Kit'', Concurrency and Computation: Practice
and Experience, 13(8-9), 2001.


\bibitem{Sto01}H. Stockinger, A. Samar, B. Allcock, I. Foster,
K. Holtman, B. Tierney. "File and Object Replication in Data Grids."
Proceedings of the Tenth International Symposium on High Performance
Distributed Computing (HPDC-10), IEEE Press, August 2001

\bibitem{Sto03}
Heinz Stockinger, Flavia Donno, Erwin Laure, Shahzad Muzaffar,
Giuseppe Andronico, Peter Kunszt, Paul Millar. ``Grid Data Management in
Action: Experience in Running and Supporting Data Management Services
in the EU DataGrid Project'', Computing in High Energy Physics (CHEP
2003), La Jolla, California, March 24 - 28, 2003.

\bibitem{Bloom}
B. Bloom ``Space/time tradeoffs in hash coding with allowable
errors'', CACM, 13(7):422-426, 1970.  

\end{thebibliography}
\end{document}